\documentclass[twocolumn,trackchanges]{aastex631}
\pdfoutput=1 %for arXiv submission
\usepackage[T1]{fontenc}
\usepackage[figure,figure*]{hypcap}
\usepackage{amsmath,amstext,bm}
\usepackage{graphicx,textcomp,fancyhdr,hyperref}
\DeclareRobustCommand{\okina}{%
\raisebox{\dimexpr\fontcharht\font`A-\height}{%
    \scalebox{0.8}{`}%
  }%
}

\newcommand{\iorb}{i_\text{orb}}

\newcommand{\iorbc}{i_{\text{orb},c}}

\newcommand{\sini}{\sin i_\star}
\newcommand{\vsini}{v\sini}

\newcommand{\Teff}{T_\text{eff}}

\newcommand{\Mearth}{M_\oplus}
\newcommand{\Rearth}{R_\oplus}
\newcommand{\Mjup}{M_J}

\newcommand{\ms}{m\,s\ensuremath{^{-1}}}
\newcommand{\kms}{km\,s\ensuremath{^{-1}}}

\begin{document}

\title{Obliquity Constraints for the Extremely Eccentric Sub-Saturn Kepler-1656~b}
\shorttitle{The Obliquity of Kepler-1656~\textnormal{b} From KPF}
\shortauthors{Rubenzahl et al.}
% \watermark{DRAFT}

\newcommand{\caltechastro}{Department of Astronomy, California Institute of Technology, Pasadena, CA 91125, USA}
\newcommand{\JPL}{Jet Propulsion Laboratory, California Institute of Technology, 4800 Oak Grove Drive, Pasadena, CA 91109}
\newcommand{\wmko}{W. M. Keck Observatory, Waimea, HI 96743, USA}
\newcommand{\UCO}{UC Observatories, University of California, Santa Cruz, CA 95064, USA}
\newcommand{\ipac}{NASA Exoplanet Science Institute/Caltech-IPAC, MC 314-6, 1200 E. California Blvd., Pasadena, CA 91125, USA}

% Authorlist
\author[0000-0003-3856-3143]{Ryan A. Rubenzahl}
\altaffiliation{NSF Graduate Research Fellow}
\affiliation{\caltechastro}

\author[0000-0001-8638-0320]{Andrew W. Howard}
\affiliation{\caltechastro}

\author[0000-0003-1312-9391]{Samuel Halverson}
\affiliation{\JPL}

\author[0000-0003-0412-9314]{Cristobal Petrovich}
\affiliation{Department of Astronomy, Indiana University, 727 East 3rd Street, Bloomington, IN 47405-7105, USA}
\affiliation{Millennium Institute for Astrophysics, Santiago, Chile}

\author[0000-0002-9751-2664]{Isabel Angelo}
\affiliation{Department of Physics \& Astronomy, University of California Los Angeles, Los Angeles, CA 90095, USA}

\author[0000-0001-7409-5688]{Gu{\dh}mundur Stef{\'a}nsson}
\affil{Anton Pannekoek Institute for Astronomy, University of Amsterdam, Science Park 904, 1098 XH Amsterdam, The Netherlands}

\author[0000-0002-8958-0683]{Fei Dai}
\affiliation{Institute for Astronomy, University of Hawai{\okina}i, 2680 Woodlawn Drive, Honolulu, HI 96822, USA}

\author[0000-0002-5812-3236]{Aaron Householder}
\affiliation{Department of Earth, Atmospheric and Planetary Sciences, Massachusetts Institute of Technology, Cambridge, MA 02139, USA}
\affil{Kavli Institute for Astrophysics and Space Research, Massachusetts Institute of Technology, Cambridge, MA 02139, USA}

\author[0000-0003-3504-5316]{Benjamin Fulton}
\affiliation{\ipac}

\author[0009-0004-4454-6053]{Steven R. Gibson}
\affil{Caltech Optical Observatories, Pasadena, CA, 91125, USA}

\author[0000-0001-8127-5775]{Arpita Roy}
\affiliation{Astrophysics \& Space Institute, Schmidt Sciences, New York, NY 10011, USA}

\author[0000-0003-3133-6837]{Abby P. Shaum}
\affiliation{\caltechastro}

\author[0000-0002-0531-1073]{Howard Isaacson}
\affiliation{501 Campbell Hall, University of California at Berkeley, Berkeley, CA 94720, USA}

 %%%%%%%%%%%%  %%%%%%%%%%%%  %%%%%%%%%%%%  

\author[0009-0008-9808-0411]{Max Brodheim}
\affiliation{\wmko}

\author[0009-0000-3624-1330]{William Deich}
\affil{\UCO}

\author[0000-0002-7648-9119]{Grant M. Hill}
\affiliation{\wmko}

\author[0000-0002-6153-3076]{Bradford Holden}
\affil{\UCO}

\author[0000-0001-8832-4488]{Daniel Huber}
\affiliation{Institute for Astronomy, University of Hawai{\okina}i, 2680 Woodlawn Drive, Honolulu, HI 96822, USA}
\affiliation{Sydney Institute for Astronomy (SIfA), School of Physics, University of Sydney, NSW 2006, Australia}

\author[0000-0003-2451-5482]{Russ R. Laher}
\affiliation{\ipac}

\author[0009-0004-0592-1850]{Kyle Lanclos}
\affiliation{\wmko}

\author[0009-0008-4293-0341]{Joel N. Payne}
\affiliation{\wmko}

\author[0000-0003-0967-2893]{Erik A. Petigura}
\affiliation{Department of Physics \& Astronomy, University of California Los Angeles, Los 
Angeles, CA 90095, USA}

\author[0000-0002-4046-987X]{Christian Schwab} % mail.chris.schwab@gmail.com
\affiliation{School of Mathematical and Physical Sciences, Macquarie University, Balaclava Road, North Ryde, NSW 2109, Australia}

\author[0000-0002-6092-8295]{Josh Walawender}
\affiliation{\wmko}

\author[0000-0002-6937-9034]{Sharon X.~Wang}
\affiliation{Department of Astronomy, Tsinghua University, Beijing 100084, People's Republic of China}

\author[0000-0002-3725-3058]{Lauren M. Weiss}
\affiliation{Department of Physics and Astronomy, University of Notre Dame, Notre Dame, IN 46556, USA}

\author[0000-0002-4265-047X]{Joshua N.\ Winn}
\affiliation{Department of Astrophysical Sciences, Princeton University, Princeton, NJ 08544, USA}

\newcommand{\PSUAA}{Department of Astronomy \& Astrophysics, 525 Davey Laboratory, Penn State, University Park, PA, 16802, USA}
\newcommand{\PSUCEHW}{Center for Exoplanets and Habitable Worlds, 525 Davey Laboratory, Penn State, University Park, PA, 16802, USA}
\newcommand{\PSETI}{Penn State Extraterrestrial Intelligence Center, 525 Davey Laboratory, Penn State, University Park, PA, 16802, USA}
\author[0000-0001-6160-5888]{Jason T.\ Wright}
\affil{\PSUAA}
\affil{\PSUCEHW}
\affil{\PSETI}

\newcommand{\lam}{35.0}
\newcommand{\elamlo}{21.6}
\newcommand{\elamhi}{14.9}

\newcommand{\vsi}{3.2}
\newcommand{\evsilo}{0.4}
\newcommand{\evsihi}{0.5}
% Abstract – no more than 250 words
% Main Text – no more than 3500 words (not including acknowledgments, appendices or other supplementary material)
% Figures and Tables – no more than 5 combined figures (each limited to 9 panels) and tables, e.g. 3 figures and 2 tables.

%%%%%%% Abstract
\begin{abstract}
The orbits of close-in exoplanets provide clues to their formation and evolutionary history. Many close-in exoplanets likely formed far out in their protoplanetary disks and migrated to their current orbits, perhaps via high-eccentricity migration (HEM), a process that can also excite obliquities. A handful of known exoplanets are perhaps caught in the act of HEM, as they are observed on highly eccentric orbits with tidal circularization timescales shorter than their ages. One such exoplanet is Kepler-1656~b, which is also the only known non-giant exoplanet ($<$100~$\Mearth$) with an extreme eccentricity ($e=0.84$). We measured the sky-projected obliquity of Kepler-1656~b by observing the Rossiter-McLaughlin effect during a transit with the Keck Planet Finder. Our data are consistent with an aligned orbit, but are also consistent with moderate misalignment with $\lambda < 50\deg$ at 95\% confidence, with the most likely solution of $\lam_{-\elamlo}^{+\elamhi}{}\deg$.
%$|\lambda| = \lam_{-\elamlo}^{+\elamhi}{}\deg$.
A low obliquity would be an unlikely outcome of most eccentricity-exciting scenarios, but we show that the properties of the outer companion in the system are consistent with the coplanar HEM mechanism. Alternatively, if the system is not relatively coplanar ($\lesssim 20\deg$ mutual inclination), Kepler-1656~b may be presently at a rare snapshot of long-lived eccentricity oscillations that do not induce migration. Kepler-1656~b is only the fourth exoplanet with $e>0.8$ to have its obliquity constrained; expanding this population will help establish the degree to which orbital misalignment accompanies migration. Future work that constrains the mutual inclinations of outer perturbers will be key for distinguishing plausible mechanisms.
\end{abstract}

% Intro
\section{Introduction} \label{6sec:intro}

High-eccentricity migration (HEM) is a leading explanation for the formation of close-in giant exoplanets, such as hot Jupiters (HJs), with orbital periods less than 10~days \citep{DawsonJohnsonReview, Rice2022}. In the HEM scenario, the giant planet forms beyond a few AU and is excited to extremely high eccentricity ($e > 0.9$) either through planet-planet scattering \citep[e.g.,]{Rasio1996} or secular interactions with an outer planetary or stellar companion \citep{Fabrycky2007, Naoz2011, Teyssandier2013}. To excite large enough eccentricities, the companion must either possess a large mutual inclination to the inner planet, in which case von-Zeipel Kozai-Lidov \citep[ZKL;][]{Kozai1962, Lidov1962, ZKL} oscillations can occur, or the companion must have an eccentric orbit, in which case higher-order eccentric Kozai-Lidov \citep[EKL;][]{Naoz2013, Naoz2013GR, Naoz2016} oscillations can yield the same result. Subsequently, tidal interactions with the star at periastron passage \citep[e.g.,][]{Rasio1996, Wu2018} circularize the orbit, causing the planet to migrate. These eccentricity-exciting mechanisms can also excite a broad range of stellar obliquities (see \citealt{AlbrechtReview} for a review),
% The HEM process can also increase the stellar obliquity, 
the angle between the host star's rotation axis and the normal to the exoplanet's orbital plane. Obliquity damping mechanisms may then come into play, as we observe HJs to be aligned around cool stars \citep[below the Kraft Break, $\lesssim 6250$~K;][]{KraftBreak} and misaligned around hot stars \citep{Winn2010hotStars, Schlaufman2010}. Cool stars are able to damp HJ obliquities faster than the age of the system either through tidal effects in their convective envelopes \citep{Albrecht2012, Lai2012, Dawson2014} or resonance locking in their radiative cores \citep{Zanazzi2024}. 

There is no reason, however, that HEM should be limited to giant planets. Many small ($< 100~\Mearth$) close-in exoplanets likely migrated from further out in their protoplanetary disks, as evidenced by their large envelope mass fraction (e.g., WASP-107~b, \citealt{Piaulet2020}) and/or highly inclined orbit \citep{AlbrechtReview, DREAMII}. Misaligned orbits could arise from the HEM process or be excited post-formation through interactions such as resonance crossing during the disk-dispersal stage \citep{Petrovich2020} or nodal precession cycles \citep{Yee18, Rubenzahl2021:WASP107}, in either case with the same outer companion that triggered HEM. Though, the census of outer companions to close-in small planets with measured obliquities is relatively incomplete. Only HAT-P-11 \citep{Yee18} and WASP-107 \citep{Piaulet2020} have fully-resolved outer companions. While these mechanisms require a large mutual inclination between the inner and outer planet, only HAT-P-11 has a mutual inclination measurement \citep[near polar;][]{Xuan2020}.

% While both HEM and nodal precession require the outer companion to have some misalignment itself, a nodally precessing inner planet will only oscillate between zero and twice the mutual inclination with the outer companion. Thus, the companion must have a $>45\deg$ orbital inclination for the inner planet to reach polar or retrograde orbits.

\begin{figure*}
    \centering
    \includegraphics[width=0.95\textwidth]{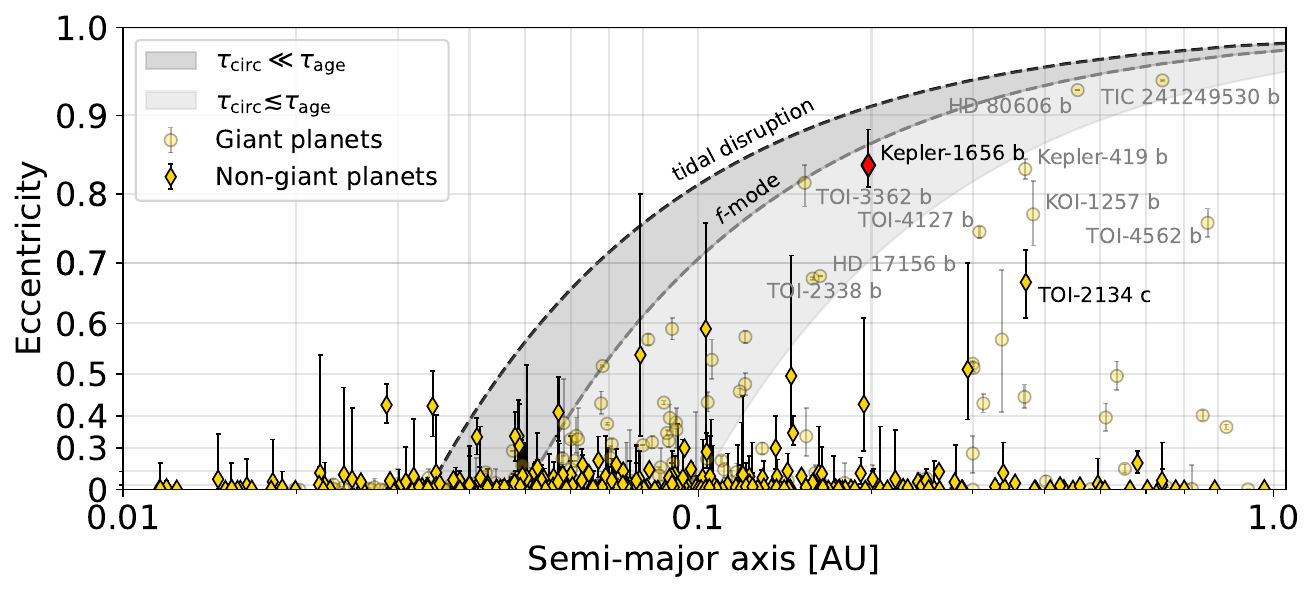}
    \caption{Eccentricity-semimajor axis diagram for transiting exoplanets. Giant planets, defined as $>$$100~\Mearth$ or $>$$8~\Rearth$ if there is no mass measurement, are plotted as faded points. Following \citet{Dong2021}, the y-axis is scaled uniformly in $e^2$, and the high-eccentricity migration track is shaded in light grey. The dashed grey line traces $a = 0.05/(1-e^2)$, corresponding to the minimum semimajor axis to excite f-mode oscillations and produce rapid orbital decay \citep{Wu2018}. Planets above this line would therefore be extremely rare. The black dashed upper boundary ($0.034/(1-e^2)$) represents the line of constant angular momentum where a 1~$M_J$, 1.3~R$_J$ exoplanet would become tidally disrupted at its closest approach to a solar mass star \citep{Dong2021}; no giant planets can persist above this boundary. The lower boundary ($0.1/(1-e^2$) corresponds to a final semimajor axis of 0.1~AU, beyond which circularization timescales become much longer than typical system ages. The boundaries are not exact as they depend on the strength and efficiency of tides in the system. Exoplanets with $e > 0.6$ are labelled; Kepler-1656~b is the only sub-Saturn firmly in the HEM track. Data are from the NASA Exoplanet Archive, accessed on 2024-04-01 \citep{neadoi}.}
    \label{6fig:ecc-sma}
\end{figure*}

If exoplanets do experience HEM, then we should expect to observe a few systems in the act of migration \citep{Socrates2012}. Approximately half a dozen known exoplanets have an eccentricity and semimajor axis such that the tidal circularization timescale is less than the age of the system (so the planet's orbit should still be circularizing), and such that the planet is not expected to be engulfed by its host star; see Figure~\ref{6fig:ecc-sma}. Several of these have obliquity measurements. Most recently, the proto-HJ TOI-3362~b \citep{Dong2021} was found to be aligned to within $3\deg$ \citep{Espinoza-Retamal2023}. This striking result indicates that perhaps some planets are able to migrate without obtaining a large obliquity. \citet{Petrovich2015CHEM} found that coplanar HEM (CHEM) can occur as a result of EKL oscillations between an outer planetary perturber (``c'') and the inner proto-HJ (``b''), provided the outer planet be relatively eccentric ($e_c>0.67$ if $e_b=0$ or both $e_b,\,e_c > 0.5$) and the planets maintain a low mutual inclination ($<20\deg$). However, any outer planetary companions in the TOI-3362 system remain undetected.

Kepler-1656~b is another such highly eccentric exoplanet \citep[$e_b = 0.84 \pm 0.01$, $48~\Mearth$, 31~day;][]{Brady2018}, and is the only known member of this class less massive than $100~\Mearth$. Kepler-1656~b is also the only highly eccentric exoplanet with a known outer planetary companion. \citet{Angelo2022} discovered Kepler-1656~c ($M_c\sin{\iorb}_{,c} = 0.4\pm0.1~\Mjup$), a giant planet in a wide ($\sim$2000~day) and  eccentric ($e_c = 0.53\pm0.05$) orbit. These authors ran a suite of dynamical integrations using the EKL formalism \citep{Naoz2016} and found that those which matched the observed system properties either rapidly ($\lesssim 1$~Gyr) circularized into a HJ-like orbit (or in a few cases crashed into the star), or achieved high eccentricity through EKL oscillations which could persist much longer than the age of the system (6.3~Gyr) without inducing tidal migration. These solutions tended to occur at high (60$\deg$--130$\deg$) mutual inclinations and would often ($\gtrsim 75\%$) excite planet b to large stellar obliquities ($>40\deg$).

% These authors ran a suite of dynamical integrations using the EKL formalism \citep{Naoz2016} assuming an in-situ, circular orbit for Kepler-1656~b, and a range of possible configurations for the perturber, Kepler-1656~c. These simulations spanned an isotropic mutual inclination distribution and resulted in a range of possible outcomes: (i) 0.8\% had planet b crash into the star, (ii) 10\% migrated and circularized into a tidally-locked orbit, (iii) 0.1\% went unstable, and (iv) in 3\% of simulations the inner planet did not migrate at all, but instead eccentricity oscillations at constant semimajor axis persisted for many Gyr with the eccentricity varying from $\sim$0 to $\sim$0.7. 85.5\% of simulations did not produce highly eccentric orbits for Kepler-1656~b, and so were discarded as possible origins. Since the time spent in the high-eccentricity stage of migration in scenario (ii) was small ($\lesssim 1$~Gyr) compared to the age of the system (6.3~Gyr), \citet{Angelo2022} concluded that scenario (iv) was most plausible. In this scenario, and in those which produced migration, the mutual inclination distribution was around 60$\deg$--130$\deg$ and planet b possessed a broad range of misaligned orbits, with a slight preference towards polar obliquities. The Kepler-1656 system is thus perhaps a proto-WASP-107-like system. Measuring its present-day stellar obliquity ($\psi$) may help to establish whether such worlds obtain large misalignments during the migration process, or if their obliquities are generated post-migration. 

In this letter, we report our measurement of the stellar obliquity of Kepler-1656~b. We present our observations of a single transit of Kepler-1656~b with the Keck Planet Finder in Section~\ref{6sec:observations}. In Section~\ref{6sec:obliquity} we modeled the Rossiter-McLaughlin \citep[RM;][]{Rossiter1924, McLaughlin1924} effect in our transit radial velocity time series and derived a projected stellar obliquity of $|\lambda| = \lam_{-\elamlo}^{+\elamhi}{}\deg$, though the data are fully consistent with an aligned orbit. We discuss the implications of this result on the dynamical history of the system and place Kepler-1656~b in the context of the broader exoplanet population in Section~\ref{6sec:discussion}, and conclude in Section~\ref{6sec:conclusion}.

\section{Obliquity Measurement}

\subsection{Observations}\label{6sec:observations}

\begin{figure*}
    \centering
    \includegraphics[height=0.38\textwidth]{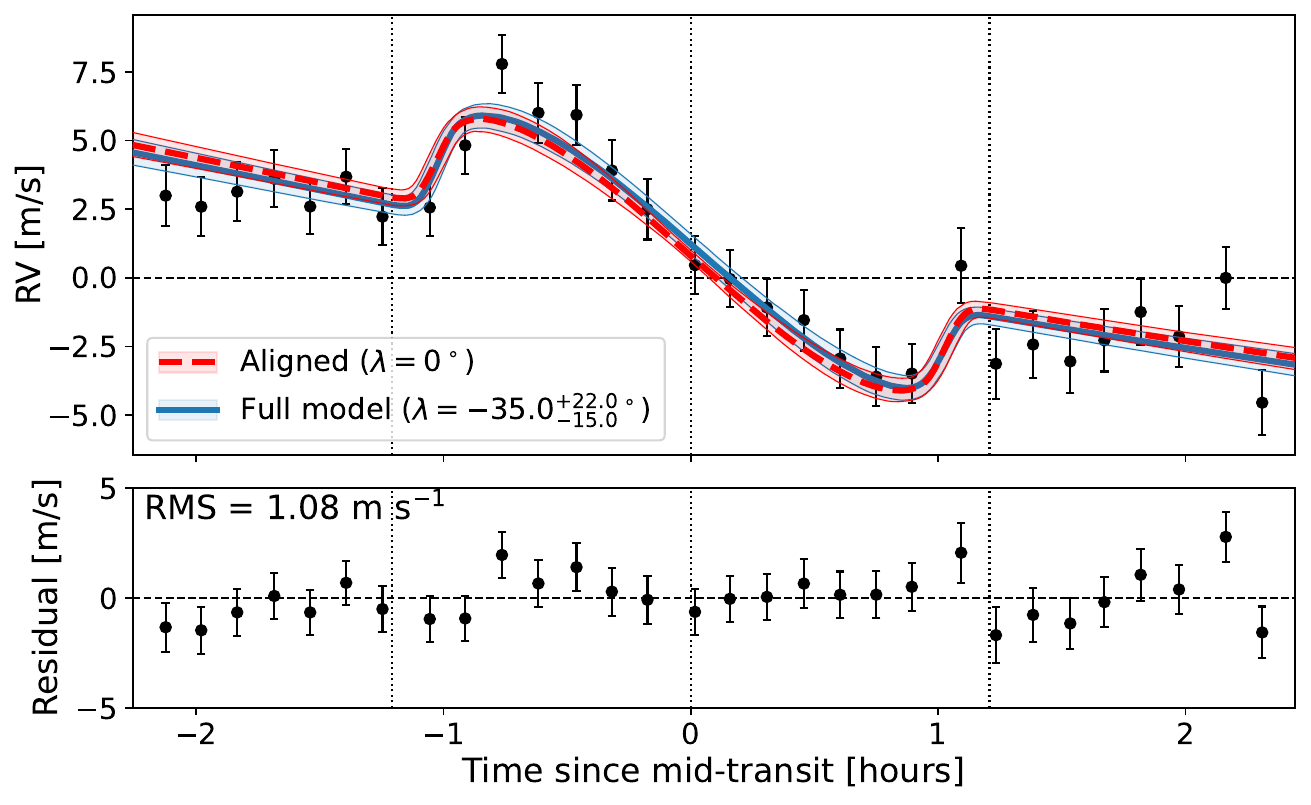}
    \includegraphics[height=0.38\textwidth]{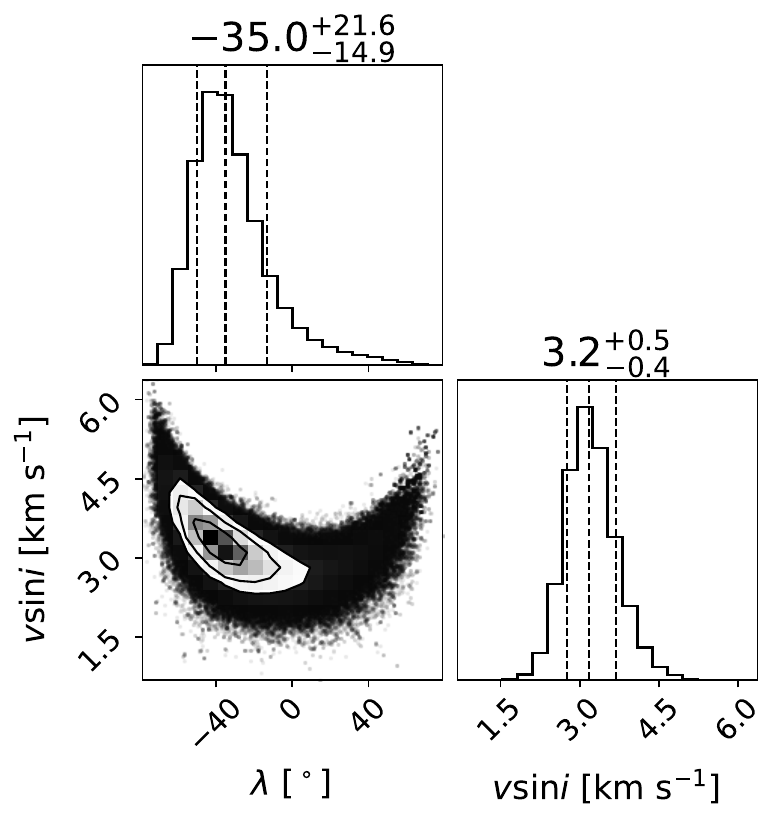}
    \caption{\textbf{Left:} The KPF RV time series, in black, during the transit of Kepler-1656~b. The blue curve shows the median RV from the posterior distribution of the full model, with the shaded band denoting the 16th--84th percentiles. The red curve shows the same for the aligned model where $\lambda$ is fixed to $0\deg$. The bottom panel shows the residuals to the median full model. \textbf{Right:} The posterior distribution for $\lambda$ and $\vsini$ for the full model. Misaligned $\lambda$ requires $\vsini > 3$~{\kms}.}
    \label{6fig:rv-rm}
\end{figure*}

We observed a single transit of Kepler-1656~b on UT June 30, 2023 with the Keck Planet Finder \citep[KPF;][]{Gibson2016, Gibson2018, Gibson2020}. We used a fixed exposure time of 480~sec to average over solar-type oscillations \citep{Brown1991, Chaplin2019} and reach a typical signal-to-noise ratio (S/N) of 100. We used the public KPF data reduction pipeline (DRP)\footnote{\href{https://github.com/Keck-DataReductionPipelines/KPF-Pipeline/}{https://github.com/Keck-DataReductionPipelines/KPF-Pipeline/}} to derive cross-correlation functions \citep[CCFs;][]{Baranne1996} using the G2 ESPRESSO weighted binary mask \citep{Pepe2002}. Orders with heavy telluric contamination are thus automatically not included in the analysis. We obtained the radial velocity (RV) as the centroid of a fitted Gaussian to the CCF. We separately extracted an RV from the green and red channels of KPF and combined the two in a weighted average, taking the weight to be proportional to the relative flux in each channel and constant in time. The resulting RV time series (shown in Figure~\ref{6fig:rv-rm}) spans a baseline from 1~hr before to 1~hr after the transit. %To track instrumental drift we interlaced hourly internal calibration frames with the etalon illuminating all three KPF fibers. The etalon RVs had typical precision $6~{\cms}$ and showed $< 1$~{\cms} drift throughout the observation baseline.

% Obliquity
\subsection{Rossiter-McLaughlin Modeling}\label{6sec:obliquity}

We fit the RV time series using \texttt{rmfit} \citep{Gummi2022}, a Python-based model for the anomalous RV produced by the RM effect based on the equations of \citet{Hirano11}. We adopted the ephemeris from \citet{Brady2018} based on their fits to the \textit{Kepler} transit light curves and adopted their fitted values as Gaussian priors for the time of conjunction ($t_c$), orbital period ($P_\text{orb}$), transit depth ($R_p/R_\ast$), scaled semimajor axis ($a/R_\ast$), orbital inclination ($\iorb$), orbital eccentricity ($e$), and argument of periastron ($\omega$). Limb darkening coefficients for the KPF bandpass ($V$) were computed with \texttt{EXOFAST} \citep{exofast}, incorporating the spectroscopic $\Teff$ ($5731\pm60$~K), [Fe/H] ($0.19\pm0.04$), and $\log(g)$ ($4.37\pm0.10$) from \citet{Brady2018}. 

There are two existing literature measurements of $\vsini$ for Kepler-1656. The California-\textit{Kepler} Survey \citep{CKSI} reported a value of $2.8\pm1.0$~{\kms} from a \texttt{SpecMatch-Synthetic} \citep{Petigura2015} analysis of a Keck/HIRES spectrum. We reanalyzed the same HIRES spectrum with \texttt{SpecMatch-Synthetic} and instead obtained an upper bound of $< 2$~{\kms}, which \texttt{SpecMatch-Synthetic} reports if the spectrum is dominated by instrument broadening. However, \citet{Masuda2022} found while analyzing $>100$ Keck/HIRES FG-type spectra that the population-level distributions obtained by applying \texttt{SpecMatch-Synthetic} to their sample were more consistent if upper-limits were instead interpreted as $< 3$~{\kms}. Kepler-1656 also appeared in the catalog of \citet{Brewer2016}, who found a smaller $\vsini$ of 1.1~{\kms} with 3.2~{\kms} macroturbulence, though this is also likely limited by instrumental broadening. As a result, we opted to place a Gaussian prior on $\vsini$ of $2.8\pm1.0$~{\kms} from the CKS result and not restrict $\vsini$ to any upper-bound.

The main free parameter in our model is the sky-projected obliquity, $\lambda$. The parameter quantifying the non-rotational line broadening, $v_\beta$, was unconstrained by the RM data, so we fixed this value to 3~{\kms} based on a 2.6~{\kms} (FWHM) instrumental broadening from KPF and the intrinsic line dispersion for $\Teff = 5731$~K from Eq. (20) in \citep{Hirano11}. Our model is a combination of a Keplerian RV signal and the RM effect, with an arbitrary offset term ($\gamma$). For the Keplerian term, we adopted the best-fit $K_b$ from \citet{Angelo2022} as a prior. Because of the low amplitude of the RM signal ($\sim$4.5~{\ms}), we also included the convective blueshift effect parameterized by $v_{CB}$ using the prescription of \citet{ShporerBrown2011}, which can contribute at the {\ms} level. We set a wide prior of $\pm10$~{\kms} on $v_{CB}$. Lastly, we include a RV jitter term ($\sigma_\text{jit}$) to account for any underestimated white noise.
% Vcb: Solar -300 m/s, K-star -200 m/s, F-star -1000 m/s

We first found the maximum a-posteriori (MAP) solution using the \texttt{PyDE} differential evolution optimizer \citep{pyde}. This solution was used as a starting point for a Markov-Chain Monte Carlo (MCMC) exploration of the posterior. We ran an \texttt{EnsembleSampler} with 100 walkers using the package \texttt{emcee} \citep{emcee}, each of which obtained 30,000 samples. We discarded the first 10\% as ``burn-in'' and checked for convergence by requiring the Gelman--Rubin statistic~\citep{Gelman2003} be $\ll 1\%$ of unity for all parameters and ensuring that the autocorrelation time was $< 2\%$ the length of the independent chains per walker \citep{Hogg2018}. The posteriors in $\lambda$ and $\vsini$ were unaffected by the inclusion of $v_{CB}$, which itself is orthogonal to the RM effect and was not detected ($v_{CB} = -560 \pm 930$~{\kms}). As a result, we fixed $v_{CB} = 0$~{\kms} in our final fit.

Our best-fitting RM model, shown in Figure~\ref{6fig:rv-rm}, has $|\lambda| = \lam_{-\elamlo}^{+\elamhi}{}\deg$. The negative $\lambda$ solutions, which the MCMC exploration finds, correspond to $\iorb < 90\deg$. However, a symmetric positive solution for $\lambda$ exists for $\iorb>90\deg$. There is also a degeneracy between $\lambda$ and $\vsini$ due to the central (low impact parameter) transit. The RM fit found $\vsini = \vsi_{-\evsilo}^{+\evsihi}$~{\kms}, which is consistent with the results from spectral broadening but is also degenerate with varying $\lambda$ (see Figure~\ref{6fig:rv-rm}, right panel). The spectral fitting results downweight RM solutions with $\vsini$ significantly greater than 3~{\kms}, i.e., larger values of $\lambda$. The full set of fitted parameters is tabulated in Table~\ref{6tab:kepler1656rmparams}.

For comparison, we also fit a model with $\lambda$ fixed to $0\deg$. This aligned model, also shown in Figure~\ref{6fig:rv-rm}, produces a fit with statistically indistinguishable goodness-of-fit metrics ($\chi^2$, BIC), though with a slightly larger $\sigma_\text{jit}$ term (see Table~\ref{6tab:kepler1656rmparams}). The aligned model yielded a lower $\vsini = 2.7\pm0.3$~{\kms}, which is slightly more consistent with the spectroscopic constraints. To summarize, neither an aligned model nor one with modest misalignment is ruled out by the data. Without any further information on $\vsini$, our RM measurement can only constrain $|\lambda| < 57\deg$ at 95\% confidence.

\startlongtable
\begin{deluxetable}{lrrr}
\centering
\tablecaption{Best-fit RM Parameters}
\label{6tab:kepler1656rmparams}
\tablehead{
  \colhead{Parameter} & 
  \colhead{Full model} & 
  \colhead{Aligned model} &
  \colhead{Unit}
}
\startdata
% \multicolumn{4}{l}{Model~Parameters} \\
$|\lambda$|             & $35.0_{-21.6}^{+14.9}$   & 0                         & $^\circ$   \\
$\vsini$              & $3.2_{-0.4}^{+0.5}$    & $2.7_{-0.3}^{+0.3}$    & {\ms}      \\
$i_\text{orb}$        & $88.9_{-0.5}^{+0.5}$    & $89.2_{-0.6}^{+0.6}$   & $^\circ$   \\
$e$                   & $0.824_{-0.014}^{+0.013}$ & $0.824_{-0.014}^{+0.013}$ &            \\
$\omega$              & $54.5_{-4.7}^{+4.7}$      & $54.7_{-4.7}^{+4.8}$      & $^\circ$   \\
$K$                   & $13.3_{-1.5}^{+1.6}$      & $13.4_{-1.6}^{+1.7}$      & {\ms}      \\
$\gamma$              & $-5.8_{-1.1}^{+1.1}$      & $-5.5_{-1.2}^{+1.1}$      & {\ms}      \\
$\sigma_\text{jit}$   & $0.51_{-0.33}^{+0.37}$    & $0.7_{-0.36}^{+0.34}$     & {\ms}      \\
\hline
$\Delta$BIC   &  0.0 & 5.0 &  \\
$\chi^2$   &  0.92 & 0.95 &  \\
\hline
\enddata
\tablenotetext{}{Posterior values display the 50$^{th}$ percentile with the upper and lower errorbars giving the difference relative to the 84$^{th}$ and 16$^{th}$ percentiles, respectively. The reduced $\chi^2$ includes the median $\sigma_\text{jit}$ added in quadrature to the measurement uncertainties.
}
\end{deluxetable}

\begin{figure*}
    \centering
    \includegraphics[width=\textwidth]{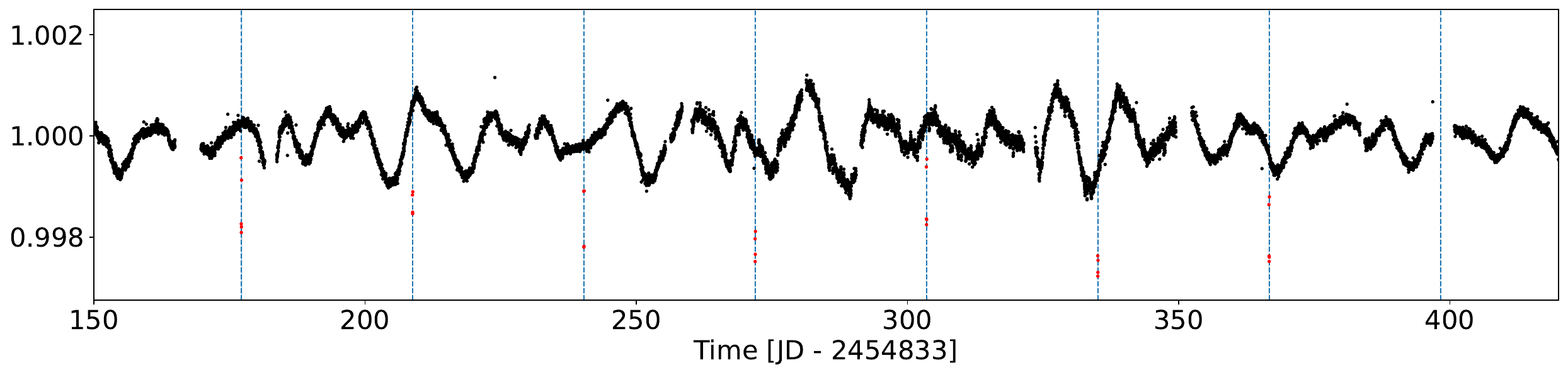}
    \includegraphics[width=0.495\textwidth]{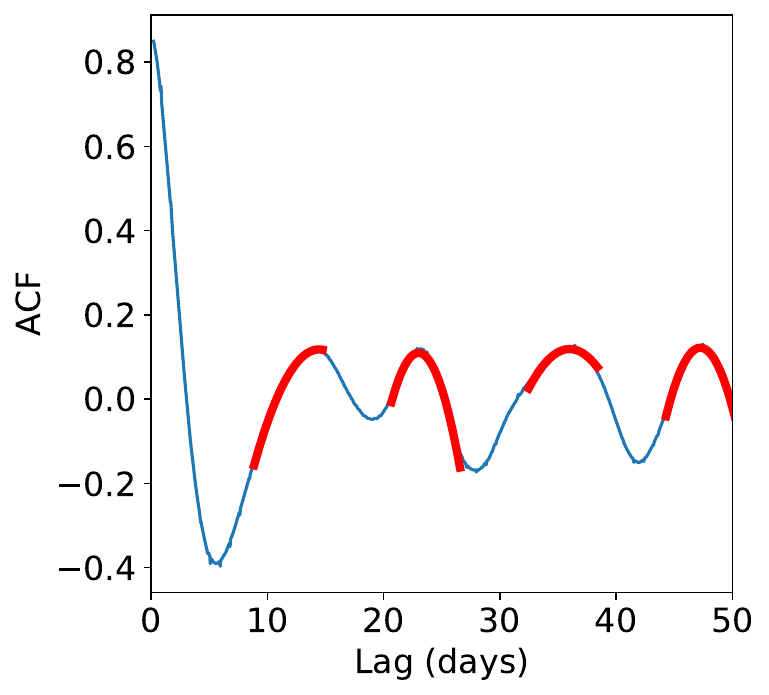}
    \includegraphics[width=0.495\textwidth]{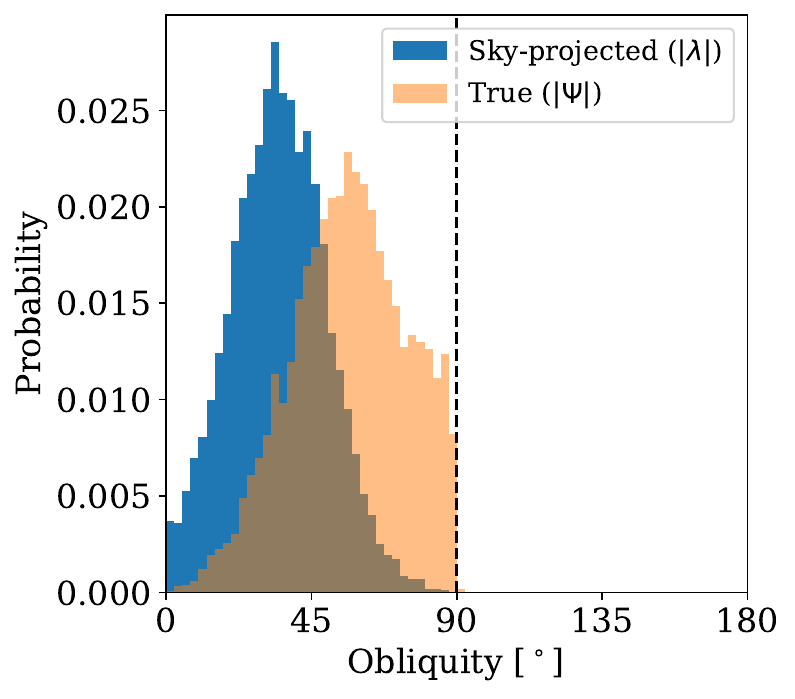}
    \caption{\textbf{Top:} Representative snippet of the \textit{Kepler} photometry. Transit windows are highlighted by the vertical blue bands, with in-transit data marked in red. \textbf{Lower left:} Autocorrelation function of the full \textit{Kepler} photometry using \texttt{spinspotter}. \textbf{Lower right:} Corresponding true obliquity distribution (orange) given the measured sky-projected obliquity (blue).}
    \label{fig:kepler_photometry}
\end{figure*}

Of note is the \textit{Kepler} photometry of Kepler-1656, shown in Figure~\ref{fig:kepler_photometry}, which exhibits significant quasiperiodic variability. An autocorrelation analysis using \texttt{spinspotter} \citep{spinspotter} detects a primary periodicity of $11.6\pm0.9$~days. If this signal corresponds to rotationally modulated starspots, then the stellar equatorial rotation speed is $4.7 \pm 1$~{\kms}. Combining this value with our constraints on $\vsini$ from the RM fit using the methodology of \citet{MasudaWinn2020}, the corresponding stellar inclination is $47\pm28 \deg$. Alltogether, this yields a constraint on the true stellar obliquity of $55_{-18}^{+20} \deg$ (Figure~\ref{fig:kepler_photometry}), which corroborates the likelihood of a misaligned Kepler-1656~b.

\section{Discussion}
\label{6sec:discussion}

\subsection{Dynamics in the Kepler-1656 system}\label{6sec:dynamics}

\begin{figure*}
    \centering
    \includegraphics[width=0.495\textwidth]{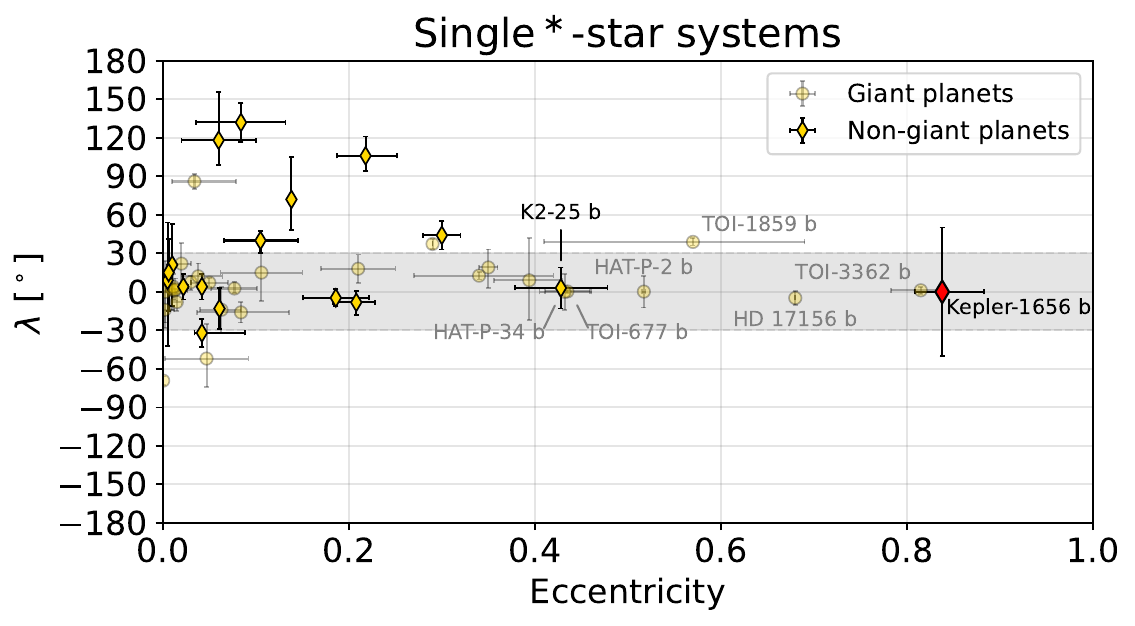}
    \includegraphics[width=0.495\textwidth]{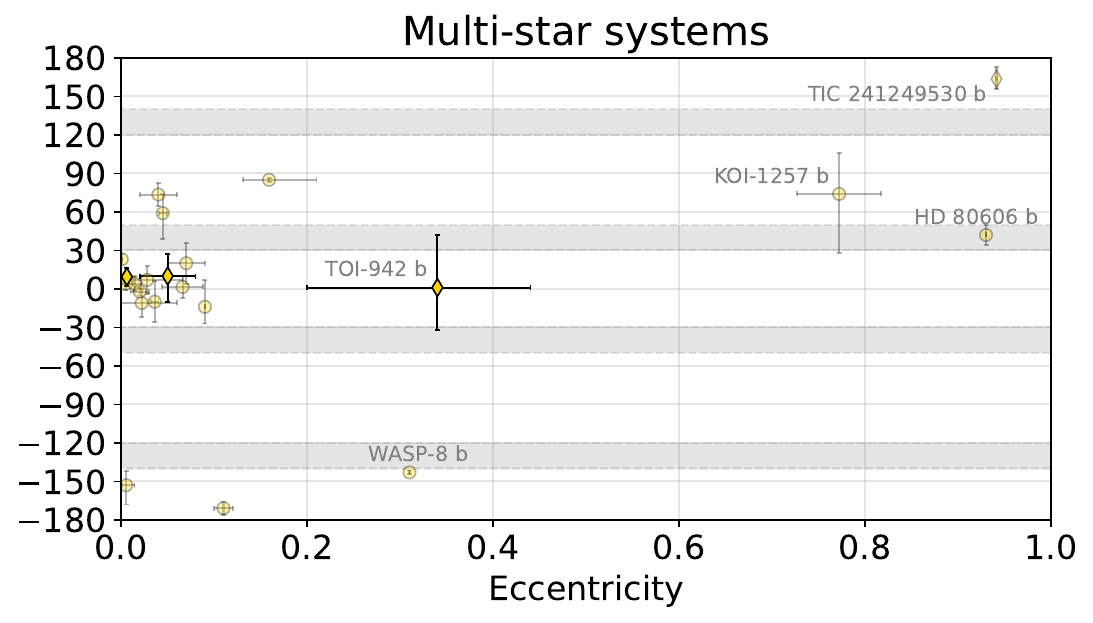}
    \caption{Projected obliquity $\lambda$ for measured systems as a function of eccentricity, for single-star systems (left) and multi-star systems. The shaded bar on the left plot covers $\pm 20\deg$, the range of obliquities for which CHEM could operate given an appropriate companion with zero stellar obliquity. The shaded bands on the multi-star plot highlight $\pm10\deg$ around the angles $\pm40\deg$ and $\pm130\deg$ corresponding to the bimodal peaks of the expected true obliquity distribution from star-planet Kozai \citep{Anderson2016}. Note that these angles refer to the true orbital inclinations, while the data points are for $\lambda$, the sky projection. The true obliquity $\psi$ is between $\lambda$ and $\text{sign}(\lambda)90\deg$. The data are the same from Figure~\ref{6fig:ecc-sma}, supplemented with updates for HD~80606~b \citep[$42\pm8\deg$;][]{Hebrard2010} and 55~Cnc~e \citep[$11^{+17}_{-20}{}\deg$][]{Zhao2023_55cnce}, and the addition of TIC~241249530~b \citep[$163.5_{-7.7}^{+9.4}{}\deg$][]{Gupta2024}), TOI-3362~b \citep[$1.2\pm2.8\deg$;][]{Espinoza-Retamal2023}, TOI 677~b \citep[$0.3\pm1.3\deg$;][]{Sedaghati2023,Hu2024}, and of course Kepler-1656~b, where we have drawn the 1$\sigma$ upper-limit of 50$\deg$ (this work).}
    \label{6fig:ecc-lam}
\end{figure*}

As a consequence of Kepler-1656~b's central transit and the uncertainty in $\vsini$, our RM dataset is consistent with an aligned orbit but cannot rule out a misaligned $\lambda$ as high as $57\deg$ at 2$\sigma$ confidence. Of the other known exoplanets in the HEM track, only TOI-3362~b and HD~80606~b have obliquity measurements. The former is aligned, with $\lambda = 1.2\pm2.8\deg$ \citep{Espinoza-Retamal2023}. Like Kepler-1656~b, TOI-3362~b orbits a single star, though it is not known if an outer giant planet exists in that system. HD~80606~b, on the other hand, is in a binary-star system and is misaligned, with $\lambda = 42\pm8\deg$ \citep{Pont2009, Hebrard2010}. Here we revisit plausible dynamics between Kepler-1656 b and c and their implications on b's obliquity.

The simulations conducted by \citet{Angelo2022} that were most consistent with Kepler-1656 b and c's orbital eccentricities and semimajor axes tended towards large mutual inclinations ($60\deg$--130$\deg$). The highly inclined companion excited $e_b$ either to the point of tidal migration or maintained long-lasting ($>$ the age of the system) eccentricity oscillations. In either case, the inner planet's obliquity tended towards misalignment; only $\sim$10\% of simulations yielded $\psi_b < 20\deg$. In the long-lasting eccentricity oscillation scenario, $\sim$75\% of simulations produced $\psi_b > 60\deg$. Simulations that migrated planet b into a tidally locked orbit were more consistent with alignment, though only $\sim$1/3rd had $\psi_b < 60\deg$.

In the low-mutual-inclination regime, an eccentric ($e_c>0.2$--$0.5$) outer companion can still excite the inner planet's eccentricity to large values $(\gtrsim0.9)$, in some cases causing the inner planet's orbit to flip 180$\deg$ from prograde to retrograde \citep{Naoz2013, Li2014}. If the eccentricity grows sufficiently large such that the periastron distance is small enough for strong tidal dissipation, the planet's orbit will shrink and circularize. \citet{Petrovich2015CHEM} showed that throughout this coplanar HEM (CHEM), the migrating inner planet maintains a low stellar obliquity ($\psi < 30\deg$) so long as the mutual inclination between the two planets is low ($\lesssim 20\deg$).

\citet{Petrovich2015CHEM} derived the initial criteria for CHEM to operate: either (i) the inner planet begins in a circular orbit, in which case the outer planet must have $e_c \gtrsim 0.67$ and $M_b/M_c(a_b/a_c)^{1/2} \lesssim 0.3$, or (ii) both planets begin eccentric ($e \gtrsim 0.5$) and $M_b/M_c(a_b/a_c)^{1/2} \lesssim 0.16$. Given the posterior distributions in mass and semimajor axis for both planets \citep{Angelo2022}, this ratio for Kepler-1656 is $M_b/M_c(a_b/a_c)^{1/2} = 0.11 \pm 0.02$, and is less than 0.16 at 99.2\% confidence. This calculation assumed $M_c\sin\iorbc$ in place of $M_c$, i.e., the planets are coplanar. The true mass of Kepler-1656~c may be larger, in which case this ratio would be smaller, still satisfying the CHEM criterion. Thus, given Kepler-1656~c's presently measured eccentricity \citep[$0.53 \pm 0.05$;][]{Angelo2022}, CHEM is a plausible explanation if either Kepler-1656~b formed with--or was able to gain--an eccentricity $> 0.5$, or if Kepler-1656~c used to be moderately more eccentric ($\gtrsim 0.67$). Such eccentricities are naturally produced by planet-planet scattering events (e.g., \citealt{Chatterjee2008}). 

\begin{figure*}
    \centering
    \includegraphics[width=\textwidth]{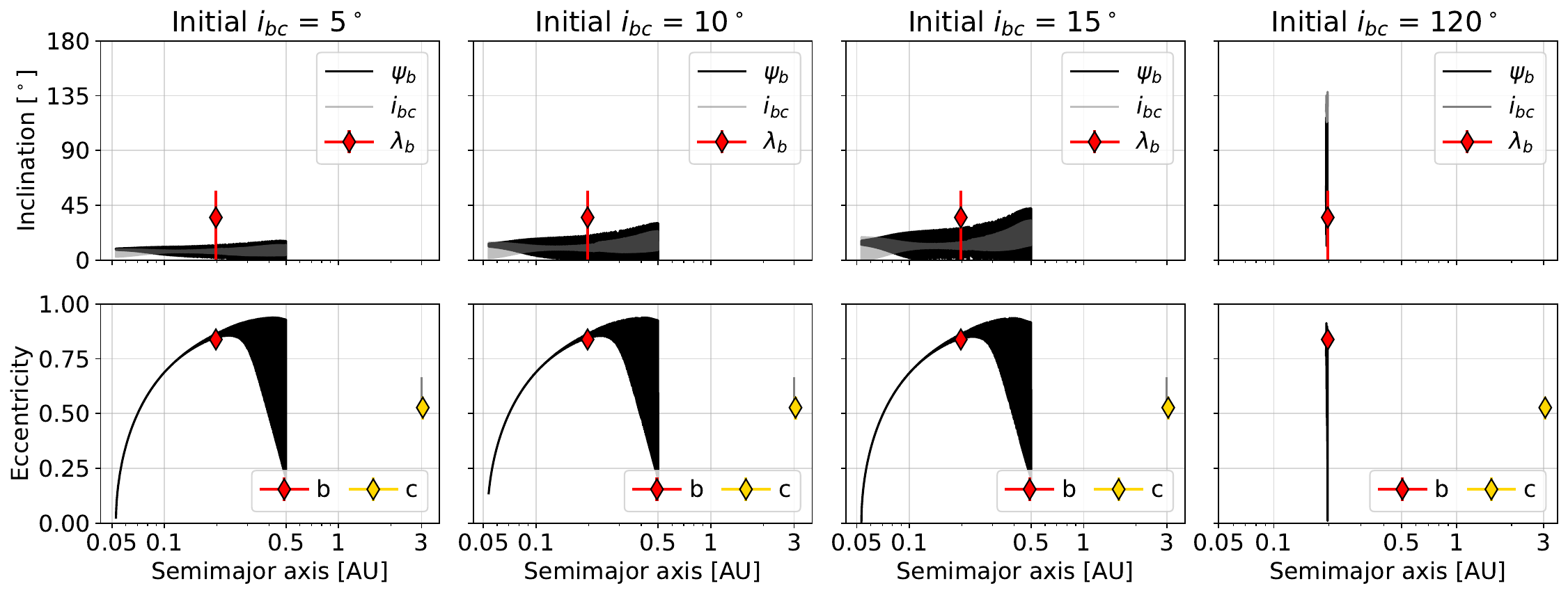}
    \caption{Three-body simulations of the orbits of Kepler-1656~b (black) and c (grey), for variable initial mutual inclinations $i_{bc}$, ($5\deg$, $10\deg$, $15\deg$, and $120\deg$ from left to right). The top row shows the evolution of the mutual inclination (grey) and the inner planet's obliquity (black), as a function of semimajor axis. The lower panel shows the eccentricity evolution of planet b (black) and c (grey). The first three ($i_{bc} \leq 15\deg$) are initialized as described in Section~\ref{6sec:dynamics}. The fourth ($i_{bc}=120\deg$) is an example of a simulation from \citet{Angelo2022} initalized \textit{in-situ} with a circular planet b. All four scenarios can reproduce the observed Kepler-1656 system (red and yellow data points). Though, $i_{bc} \leq 15\deg$ requires starting planet b at a more distant orbit that subsequently migrates through its present-day location, before circularizing in $\sim$100\,Myr; larger mutual inclinations excite large eccentricities without triggering migration, with brief excursions to low obliquity over many $\sim$Gyr. The measured projected obliquity is plotted for the positive-$\lambda$ scenario (to match $\psi_b>0$) in the top panel at $35\deg$ with errorbars covering $0\deg$--$50\deg$.}
    \label{6fig:chem}
\end{figure*}

We integrated the three-body equations of motion expanded to octupole order \citep{Ford2000} as described in the appendix of \citet{Petrovich2015Kozai}, given the measured planet masses of \citet{Angelo2022}. We initialized planet b with an eccentricity of 0.2 at semimajor axis 0.5~AU and planet c with an eccentricity of 0.67 at its present-day semimajor axis of 3~AU. We included tidal dissipation in planet b parameterized with a 0.01~yr viscous timescale\footnote{For simplicity, we have adopted a much lower viscous timescale than \citet{Angelo2022} of 1.5 yr, allowing significant migration within $\sim 10^8$ yr. For reference, a viscous timescale of 0.01~yr (1.5~yr) is equivalent to setting a tidal quality factor of $Q_b\sim10^4$ ($Q_b\sim10^6$) at the planet's current location.} and Love number 0.25. We ran three simulations with the inner planet aligned at $0\deg$ obliquity but with the outer planet at $5\deg$, $10\deg$, and $15\deg$ mutual inclination. We found in all three cases the inner planet's eccentricity became excited up to 0.94 and underwent oscillations for $\sim$100~Myr before tidal effects quenched oscillations and triggered planet b's migration into a HJ-like orbit. We stress that while the migration phase of the simulation agrees with the data, it is relatively short-lived ($<100$~Myr) and thus has a low probability of being observed. We conclude that CHEM could be operating in the Kepler-1656 system, but the data remain consistent with a non-migrating in-situ formed planet b being observed at a snapshot of high-eccentricity oscillations, as suggested by \citet{Angelo2022}. The key to distinguishing these scenarios is the planet-perturber mutual inclination. Unfortunately, at $186\pm0.5$~pc \citep{gaiadr3} Kepler-1656 (V=11.6) is too far for such a measurement with Gaia astrometry. The expected astrometric signal from Kepler-1656~c is just $5.2~\mu$as, but at $G=11$ we can expect a single-epoch precision of $34.2~\mu$as with Gaia \citep{Perryman2014}.

\subsection{Kepler-1656 in context}

\citet{Espinoza-Retamal2023} noted a dichotomy in eccentric ($e \gtrsim 0.3$) planet obliquities; those in single-star\footnote{$^\ast$The presence of stellar companions is not homogenously constrained. For simplicity, here we define ``single star'' to be those in the Exoplanet Archive with \texttt{sy\_snum==1}.} systems tend to be aligned, while those in multi-star systems tend to be misaligned. We plot the updated $\lambda$--$e$ diagram in Figure~\ref{6fig:ecc-lam}. For single-star systems, the next most eccentric sub-Saturn ($<100~\Mearth$) with a measured obliquity is K2-25~b \citep[$e=	0.43\pm0.05$, $\lambda = 3.0 \pm 16.0\deg$;][]{Stefansson2020k225}. Kepler-1656~b and HAT-P-2~b \citep{deBeurs2023}, a HJ, are the only highly eccentric ($e > 0.3$) exoplanets with fully measured outer planetary companions.
% Of the known exoplanets with ($e > 0.4$), those which have obliquity measurements include HD~80606~b \citep[$42\pm8\deg$;][]{Pont2009, Hebrard2010}, 
% Kepler-420~b \citep[KOI-1257, $74^{+32}_{-46}{}\deg$;][]{Santerne2014}, 
% and TIC~241249530~b ($160\pm8\deg$; Gupta, A. et al. in review), 
% which orbit stars in multi-star systems, 
% and TOI 677~b \citep[$0.3\pm1.3\deg$;][]{Sedaghati2023,Hu2024}, 
% HAT-P-34~b \citep[$0\pm14\deg$;][]{Albrecht2012}, 
% % TOI-2025~b \citep[$9.0^{+33.0}_{-31.0}{}\deg$;][]{Knudstrup2022}, 
% K2-25~b \citep[$17^{+11}_{-8}\deg$;][]{Stefansson2020k225},
% HD 17156~b \citep[$-4.8\pm5.3$;][]{Barbieri2009}, 
% and TOI-3362~b \citep[$1.2\pm2.8\deg$;][]{Espinoza-Retamal2023}, which orbit single-star systems. Intriguingly, the former category (eccentric planets in binary systems) are all misaligned, while the latter (eccentric planets in single-star systems) are all consistent with aligned orbits, often to within just a few degrees. Figure~\ref{6fig:ecc-lam} shows these two populations as of this writing.
% TOI 2025 has a long term RV trend consistent with a giant companion t ~2000 days and 70 jupiter masses. also Rodriguez et al. 2023 have e=0.39\pm0.03.

If close-in exoplanets, large and small alike, form primarily from HEM, Figure~\ref{6fig:ecc-lam} suggests that the identity of the perturber plays the key role in determining the obliquity. In multi-star systems, exoplanets with large obliquities span a wide range of mutual inclinations with their outer stellar companions \citep{Behmard2022,Rice2024}. Though, there is an overabundance of edge-on binary orbits for systems hosting transiting exoplanets compared to field binaries, suggestive of a tendency towards mutual alignment \citep{Dupuy2022, Rice2024}. It is still a small sample size, but the four oblique and eccentric exoplanets in multi-star systems (see Figure~\ref{6fig:ecc-lam}) have obliquities near those expected from star-planet Kozai \citep[e.g.][]{Anderson2016}, which requires mutual inclinations $> 39.2\deg$ \citep{Kozai1962, Lidov1962, Naoz2016}. \citet{Rice2024} calculated the ``linear motion parameter'' \citep[$\gamma$;][]{Tokovinin2015}, for WASP-8~B ($4.0 \pm 0.5\deg$) and HD 80606~B ($174.3 \pm 0.3\deg$). Both are consistent with edge-on orbits and perhaps mutual alignment, which in addition to the wide separation for HD~80606~B would make Kozai less likely. KOI-1257 is unresolved by \textit{Gaia}, preventing an astrometric detection. However, computing $\gamma$ for TIC 241249530 from its \textit{Gaia} DR3 astrometry \citep{gaiadr3} yields $85.9 \pm 36.0\deg$, which could indicate a near face-on orbit and thus a significant mutual misalignment. Interpreting inclinations from the $\gamma$ parameter alone is still poorly constrained and degenerate with eccentricity. Longer baseline multi-epoch astrometry is needed to constrain individual systems.
% # Star          gamma (my calc)
% # HD80606       174.78 +/- 0.36 deg
% # KOI1257       32.88 +/- 72.49 deg
% # TIC241249530  85.86 +/- 36.05 deg
% # WASP8         5.94 +/- 0.53 deg
% # Star          gamma (Rice+2024)
% # HD80606       174.3 +/- 0.3 deg
% # KOI1257       blended
% # TIC241249530  ---
% # WASP8         4.0 +/- 0.5 deg

Exoplanets in single-star systems, by definition, must instead be perturbed by outer planetary companions. Cold Jupiter companions in systems with inner small exoplanets show a tendency towards coplanarity \citep{Masuda2020}. This would preclude classical Kozai oscillations, though eccentric coplanar outer companions could still excite large eccentricities via the EKL mechanism \citep{Naoz2016}. In such systems, CHEM could produce close-in but aligned planets. On the other hand, large mutual inclinations have been observed in the $\pi$~Men and HAT-P-11 systems. Both have large (polar) mutual inclinations between inner and outer planet \citep[as evidenced by \textit{Gaia} astrometry;][]{Xuan2020}. In such systems, ongoing nodal precession will make it more likely than not to observe the planet in a misaligned orbit \citep{Becker2017}. Accordingly, $\pi$~Men~c is slightly misaligned \citep[$\psi = 26.9^{+5.8}_{-4.7}\deg$;][]{Hodzic2020} and HAT-P-11~b is near-polar \citep[$\lambda = 106_{-12}^{+15}{}\deg$;][]{Winn2010hatp11, SanchisOjeda2011}. So while there may not be a mutual inclination requirement for exciting eccentricities, and consequently triggering migration, the obliquity of the inner planet is likely still dependent. For the closest systems ($< 60$--100~pc), the full astrometric timeseries in the upcoming Gaia DR4 will enable constraints on the outer planet's inclination \citep{Espinoza-Retamal2023b}.

\section{Conclusions}
\label{6sec:conclusion}
We measured the stellar obliquity of Kepler-1656~b from the Rossiter-McLaughlin anomaly observed with the Keck Planet Finder. We found the orbit to be consistent with alignment, but could not rule out misalignments up to $57\deg$ at 2$\sigma$ confidence. Kepler-1656~b is one of four exoplanets to have an obliquity measurement that lives in the HEM track of the $e-a$ diagram. Two of these are in multi-star systems and are misaligned, while TOI-3362~b orbits a single star and is aligned. The mutual inclination of the perturber likely plays a leading role in determining the migration process and subsequently the obliquity of the migrating/migrated planet.

Since obliquity damping is less efficient for small planets \citep[$\tau_\psi \propto (M_p/M_\ast)^{-2}$;][]{Hut1981}, the obliquity distribution of small planets offers a more pristine view of post-migration obliquities, analogous to HJs around hot stars. There is a growing population within the $<100~\Mearth$ regime of polar orbits \citep{DREAMII} which has been noted for the HJs \citep{Albrecht2022} but is not yet statistically robust in that sample \citep{Dong2023, Siegel2023}. Kepler-1656~b represents a rare example of such a proto-hot-Neptune/Saturn that could be in the act of migrating, kick-started by its outer planetary companion. If obliquities are not excited in conjunction with HEM, then post-migration dynamics must be important for exciting the broad obliquity distribution we observe today.

% Have to do section title to get it to appear
\section{Acknowledgements}

Some of the data presented herein were obtained at Keck Observatory, which is a private 501(c)3 non-profit organization operated as a scientific partnership among the California Institute of Technology, the University of California, and the National Aeronautics and Space Administration. The Observatory was made possible by the generous financial support of the W. M. Keck Foundation. Keck Observatory occupies the summit of Maunakea, a place of significant ecological, cultural, and spiritual importance within the indigenous Hawaiian community. We understand and embrace our accountability to Maunakea and the indigenous Hawaiian community, and commit to our role in long-term mutual stewardship. We are most fortunate to have the opportunity to conduct observations from Maunakea. 

R.A.R.\ acknowledges support from the National Science Foundation through the Graduate Research Fellowship Program (DGE 1745301). A.W.H.\ acknowledges funding support from NASA award 80NSSC24K0161 and the JPL President's and Director's Research and Develop Fund. CP acknowledges support from ANID BASAL project FB210003, FONDECYT Regular grant 1210425, CASSACA grant CCJRF2105, and ANID+REC Convocatoria Nacional subvencion a la instalacion en la Academia convocatoria 2020 PAI77200076. D.H. acknowledges support from the Alfred P. Sloan Foundation, the National Aeronautics and Space Administration (80NSSC21K0652), and the Australian Research Council (FT200100871).

This research was carried out, in part, at the Jet Propulsion Laboratory and the California Institute of Technology under a contract with the National Aeronautics and Space Administration and funded through the President’s and Director’s Research \& Development Fund Program.

\facility{Keck:I (KPF) \citep{Gibson2020}}

\software{
\texttt{astropy}    \citep{astropy},
\texttt{corner}     \citep{corner},
\texttt{emcee}      \citep{emcee},
\texttt{matplotlib} \citep{matplotlib},
\texttt{numpy}      \citep{numpy},
\texttt{pandas}     \citep{pandas},
\texttt{PyDE}       \citep{pyde},
\texttt{rmfit}      \citep{Gummi2022},
\texttt{scipy}      \citep{scipy},
\texttt{spinspotter}\citep{spinspotter},
% \texttt{SpecMatch-Emp} \citep{smemp}, 
% \texttt{SpecMatch-Synth} \citep{Petigura2015},
}

\bibliography{references}
\bibliographystyle{aasjournal}


\end{document}